\documentclass[lettersize,journal]{IEEEtran}
\usepackage{amsmath,amsfonts}
\usepackage{algorithmic}
\usepackage{algorithm}
\usepackage{array}
\usepackage[caption=false,font=normalsize,labelfont=sf,textfont=sf]{subfig}
\usepackage{textcomp}
\usepackage{stfloats}
\usepackage{url}
\usepackage{verbatim}
\usepackage{graphicx}
\usepackage{cite}
\usepackage{hyperref}
\usepackage{color, xcolor}

\begin{document}

\title{Resource Allocation Driven by Large Models in Future Semantic-Aware Networks}

\author{Haijun Zhang,~\IEEEmembership{Fellow,~IEEE}, Jiaxin Ni, Zijun Wu, Xiangnan Liu,~\IEEEmembership{Member,~IEEE}, and V. C. M. Leung,~\IEEEmembership{Life Fellow,~IEEE}             
\thanks{Haijun Zhang, Jiaxin Ni and Zijun Wu are with Beijing Engineering and Technology Research Center for Convergence Networks and Ubiquitous Services, University of Science and Technology Beijing, Beijing, China, 10083 (email: zhanghaijun@ustb.edu.cn, M202310621@xs.ustb.edu.cn, wuzijun@xs.ustb.edu.cn).}
\thanks{Xiangnan Liu is with the School of Electrical Engineering and Computer Science, KTH Royal Institute of Technology, 11428 Stockholm, Sweden (e-mail: xiangliu@kth.se).}
\thanks{V. C. M. Leung is with the College of Computer Science and Software Engineering, Shenzhen University, Shenzhen 51806, China, and also with the Department of Electrical and Computer Engineering, The University of British Columbia (UBC), Vancouver, British Columbia V6T 1Z4, Canada (email:vleung@ieee.org).}
}




\maketitle

\begin{abstract}
Large model has emerged as a key enabler for the popularity of future networked intelligent applications. However, the surge of data traffic brought by intelligent applications puts pressure on the resource utilization and energy consumption of the future networks. With efficient content understanding capabilities, semantic communication holds significant potential for reducing data transmission in intelligent applications. In this article, resource allocation driven by large models in semantic-aware networks is investigated. Specifically, a semantic-aware communication network architecture based on scene graph models and multimodal pre-trained models is designed to achieve efficient data transmission. On the basis of the proposed network architecture, an intelligent resource allocation scheme in semantic-aware network is proposed to further enhance resource utilization efficiency. In the resource allocation scheme, the semantic transmission quality is adopted as an evaluation metric and the impact of wireless channel fading on semantic transmission is analyzed. To maximize the semantic transmission quality for multiple users, a diffusion model-based decision-making scheme is designed to address the power allocation problem in semantic-aware networks. Simulation results demonstrate that the proposed large-model-driven network architecture and resource allocation scheme achieve high-quality semantic transmission.
\end{abstract}


\section{Introduction}
\IEEEPARstart {R}{ecent} breakthroughs in deep learning have greatly advanced the development of large models in the fields of multimodal data and natural language processing \cite{ref1}. With extensive training datasets and large-scale model parameters, large models exhibit remarkable adaptability to various intelligent applications. Currently, large models are at an unprecedented stage of development and driving leaps in productivity and profound paradigm shifts across various industries, such as industrial internet and autopilot. Future mobile networks face increasing challenges due to the growing complexity of user demands, especially when it comes to providing reliable, high-quality communication with minimal delays across different areas. Large models exhibit significant potential in facilitating the intelligent adaptation of future mobile networks to these emergent data-intensive application needs.

With the continuous emergence of data-intensive applications such as augmented reality/virtual reality (AR/VR) and the metaverse, the data traffic in future mobile networks is expected to grow exponentially, which brings immense pressure on wireless communication systems \cite{ref2}. The traditional communication paradigm based on Shannon's theory is approaching its transmission theoretical limits and cannot satisfy the increasingly complex, diverse, and intelligent information transmission demands. Semantic communication has been widespread concerned from both academia and industry due to its efficient information processing and powerful data compression capabilities. Semantic communication utilizes advanced artificial intelligence techniques for semantic feature extraction and data compression of raw data. By transmitting the most relevant semantic information for the task and eliminating the transmission of redundant information, semantic communication significantly reduces the amount of data transmitted.

Semantic communication can be well integrated with intelligent applications focused on semantics rather than data bit distortion in future mobile networks, enabling more intelligent and adaptive communication services. Through large-scale pre-training, large models demonstrate powerful capabilities in semantic understanding, multimodal processing, and knowledge learning, providing unprecedented computational power and intelligent support for smart networks. By leveraging significant advantages in semantic understanding and generation, large models can achieve feature extraction and efficient resource utilization in semantic communication.

There are some studies focusing on large-model-driven semantic communication, but they mainly investigate encoding schemes \cite{ref3} and network structures \cite{ref4}. The authors of \cite{ref3} considered using ResNet as the main architecture for the semantic network encoder and achieved task-adaptive semantic compression. In \cite{ref4}, the authors designed a transformer-based framework to unify the semantic network structure for different tasks, where image reconstruction, machine translation, and visual question answering were considered. The authors in \cite{ref5} investigated a large-model-based semantic importance-aware communication scheme, by utilizing ChatGPT and BERT to quantify the semantic importance of data frames. And power allocation was performed based on this quantified semantic importance, addressing the limitations of existing semantic understanding capabilities. For image transmission tasks, scene graph generation models were utilized to obtain textual semantic feature representations of image data~\cite{ref6,ref7}. Compared to transmitting image data directly, transmitting natural language representations can significantly reduce communication overhead and alleviate the scarcity of spectrum resources. 

\begin{figure*}[!t]	
	\centering	
	\includegraphics[width=6in]{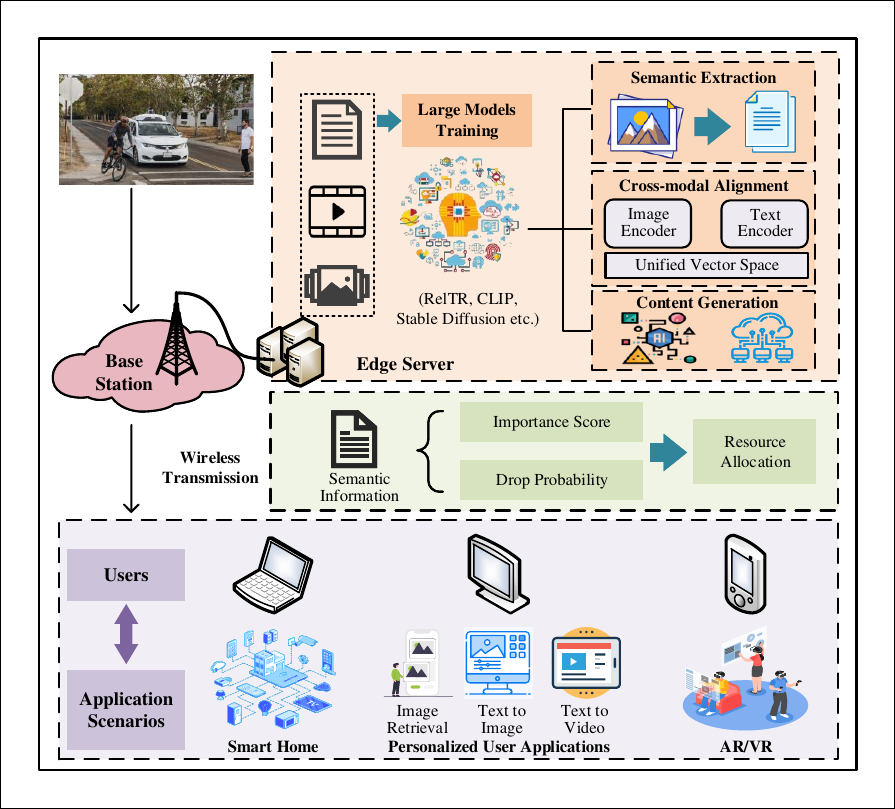}	
	\caption{The network architecture of large-model-driven semantic network.}	
	\label{fig1}
\end{figure*}

It is crucial to design an efficient large-model-driven wireless resource allocation mechanism while achieving optimal semantic information acquisition. Semantic spectrum efficiency \cite{ref8}, quality of experience \cite{ref9} composed of the semantic rate and the semantic fidelity, and task-oriented semantic spectrum efficiency \cite{ref10} were designed to measure semantic resource utilization. Meanwhile, generative diffusion large model \cite{ref11} has tremendous potential in modeling complex data distributions and generating high-quality samples. Given the superior decision-making capabilities of diffusion models, the deep integration of diffusion models with wireless communication networks has emerged as an inevitable trend. In this article, considering the goals of intelligent tasks, the quality of transmitted content, and the performance of wireless fading channels, a novel metric for semantic transmission quality is designed. Furthermore, the resource allocation problem in intelligent semantic networks is addressed by large-model-driven semantic importance awareness and diffusion-model-driven power resource allocation. The main contributions of this article can be summarized as follows:
\begin{itemize}
\item[$\bullet$] A semantic communication framework based on large models is proposed. Specifically, the proposed architecture utilizes scene graph generation models to obtain textual semantic triplets of image data, and further uses the Contrastive Language-Image Pre-Training (CLIP) \cite{ref12} model to quantify the importance of these triplets. 
\item[$\bullet$] A novel metric consisting of semantic importance and semantic triplet drop probability has been defined. This metric ensures that critical image semantic information can be transmitted to users with high quality under unstable channel conditions.
\item[$\bullet$] On the premise of the established metric for evaluating image semantic transmission, an adaptive power allocation scheme based on diffusion model is adopted to maximize the transmission quality of large-model-driven semantic communication networks.
\end{itemize}

The remainder of this article is arranged as follows. The proposed semantic communication network framework driven by large models is described in Section II. Section III introduces the proposed diffusion-model-based resource allocation in semantic-aware network. In Section IV, simulation results are presented and discussed. Subsequently, several of open issues and challenges are detailed in Section V. Finally, the conclusion is drawn in Section VI.

\begin{figure*}[!t]	
	\centering	
	\includegraphics[width=7.16in]{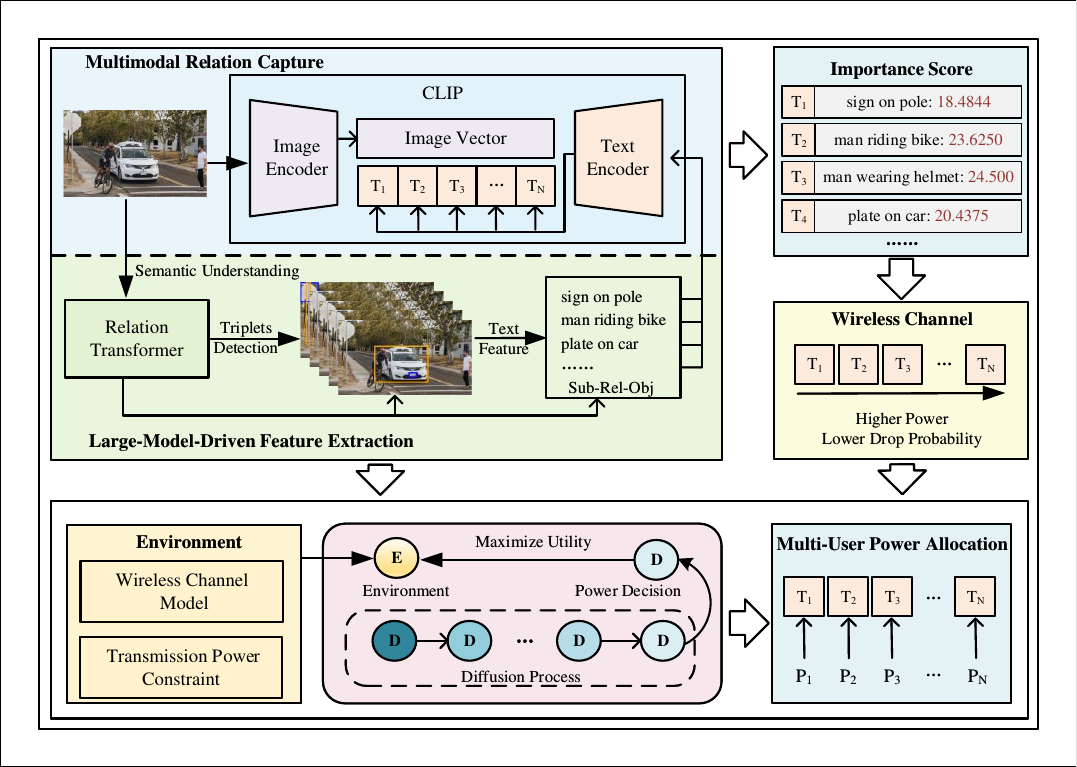}	
	\caption{The procedure of multi-user power allocation in large-model-driven network.}	
	\label{fig2}
\end{figure*}

\section{Large-model-driven semantic network}
The explainable semantic representation capabilities of scene graph generation models and the multimodal generation abilities of large models provide new insights for building more intelligent and efficient semantic networks. The architecture of large-model-driven semantic communication network is depicted in Fig. \ref{fig1}. Consider a cellular network consisting of a base station (BS) with an edge server and $U$ users, where all users download image data from the BS. The edge server enabled by large models can achieve semantic extraction, cross-modal semantic alignment, and content generation. Specifically, the edge server uses the scene graph generation model to extract the semantic information from images, resulting in a series of textual semantic feature triplets (subject-relation-object). Then, these extracted semantic triplets are compressed, encoded, and modulated before being transmitted to the users through the wireless channel. Upon receiving the textual image semantic features, users utilize the text for downstream intelligent tasks. Compared to existing semantic communication architectures for image transmission, the proposed large model-driven semantic network employs conventional bit-based data transmission, enhancing its compatibility with current wireless communication networks. This article considers the semantic communication process in two stages~\cite{ref6}: semantic information extraction and semantic information transmission.

The Relation Transformer (RelTR), as a large model designed for semantic understanding tasks, can efficiently capture the content of image data and achieve semantic alignment while performing image-to-text modality conversion. In the proposed architecture, the edge server uses RelTR model to convert images into text-based graph structures. Different from the two-step semantic information extraction method in~\cite{ref6}, a one-stage method is used to directly predict the scene graph based on the visual appearance of the image, achieving faster inference with fewer parameters. Specifically, the pre-trained RelTR model is based on an encoder-decoder architecture. The encoder infers the visual feature context of the image. At the same time, the decoder uses attention mechanisms and coupled subject and object queries to infer a set of subject-relation-object triplets \cite{ref13}. For example, a semantic triplet in Fig.~\ref{fig2} is ([“sign”], [“on”], [“pole”]), where [“sign”] is the subject, [“on”] is the relation, and [“pole”] is the object. The semantic features of an image can be represented by multiple semantic triplets, whose transmission requires fewer channel resources and provides efficient data storage for users. Additionally, the importance of transmitting semantic triplets is further quantified and transmit power is allocated accordingly to ensure the quality of important information transmission.

Furthermore, the transmission of image semantic triplets in wireless channels faces multiple challenges. Due to the continuous dynamic changes in the wireless communication environment, system transmission performance is inevitably constrained by factors such as multipath fading and interference. This directly results in the inability of the user receiver to achieve lossless reception of image semantic triplets. To address this challenge and ensure high-quality semantic communication in dynamic wireless environments, the concept of triplet drop probability is introduced, which precisely quantifies the specific impact of the wireless channel environment on the transmission of semantic triplets. In the proposed framework, time division multiplexing is utilized. Successfully and accurately transmitted semantic triplets to the user receiver can not only directly enhance the user's understanding of image content and assist in completing image retrieval tasks, but also serve as key textual prompts to drive the regeneration of similar images, thereby supporting diverse applications in intelligent mobile networks. For example, on the receiver end, generative large models (Stable Diffusion, DALL·E etc.) utilize the received semantic triplets for intelligent content generation, such as text-to-image or text-to-video, to provide dedicated application services for users. Additionally, the proposed framework can be easily extended to other related intelligent tasks in the multimodal domain, such as visual question answering and image-text generation.

\begin{figure}
        \centering	
	\includegraphics[width=3.5in]{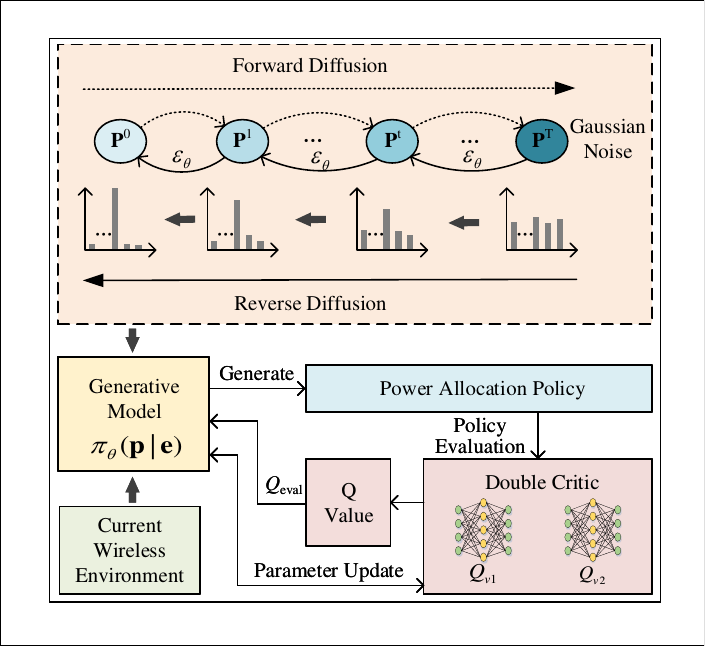}	
	\caption{The design principles of diffusion model.}	
	\label{fig3}
\end{figure}

\section{Diffusion-model-driven resource allocation in semantic network}
The rapid development of large models has opened new avenues to alleviate the data transmission challenges brought by the proliferation of data-intensive intelligent applications. Indeed, many researchers are focusing on the development of semantic communication technologies enabled by large models, particularly in the areas of encoding strategy optimization and network architecture design. However, in building large-model-driven semantic networks, what should be considered are more than a more powerful data compression and a deeper semantic understanding, but also a more effective resource allocation strategy. Considering resource scarcity and power consumption constraints, it is essential to reasonably allocate limited resources to ensure the efficient and intelligent operation of large-model-driven networks. Therefore, this article mainly focuses on the optimization of power allocation in large-model-driven wireless networks to achieve higher semantic communication efficiency. Additionally, considering the dynamic nature of wireless transmission channels and the requirements for triplet text-image similarity, the semantic transmission quality metric is established. The procedure of power resource allocation is shown in Fig. \ref{fig2}. The importance of transmitted semantic triplets is quantified using the CLIP model, and the triplet drop probability is determined based on transmit power decisions to analyze semantic transmission quality. It is noteworthy that the transmit power decision scheme of the BS is generated based on a diffusion model. 

\subsection{Semantic Transmission Quality Analysis}
Traditional communication technologies focus on the similarity between the semantic information extracted by large models and their original images, ignoring the impact of wireless channels on the transmission quality of semantic information. To more accurately evaluate the transmission efficiency of semantic communication networks, a metric for image semantic transmission is proposed. Specifically, the image semantic transmission metric consists of two components: semantic importance and semantic triplet drop probability.

 \textbf{Semantic Importance} $I_{s}$: To maximize the efficiency of semantic communication, it is necessary to quantify the importance of extracted semantic triplets and allocate transmit power accordingly to ensure the quality of important information transmission. Semantic importance characterizes the subjective user experience and is defined as the similarity between the extracted semantic triplets and the original image. Semantic importance measures the semantic relevance of the information to the original image at the semantic level. By using the pre-trained large model CLIP \cite{ref12} as an efficient vectorization tool, the relationship between textual semantic information and the image is established. Specifically, the text and image are mapped into a unified vector space by the text encoder and the image encoder, respectively, resulting in text semantic vectors and image vectors. Furthermore, the cosine of the angle between the image vector and the corresponding semantic text vector is calculated to obtain the cosine similarity score, which is used as the metric for the importance of the extracted semantic information.

\textbf{Semantic Triplet Drop Probability} $P_{d}$: As an effective metric for assessing the reliability of triplet transmission over wireless channels, semantic triplet drop probability is defined as the probability that the number of erroneous bits in a triplet exceeds the error correction capability threshold. For simplicity, assume all triplets are encoded with a fixed bit length $L_{T}$  and at the user end, the bit error correction code can correct up to $L_{E}$ erroneous bits. Once the number of erroneous bits exceeds the correction capability of the error correction code, the transmission of that triplet is considered a failure, resulting in a drop. The triplet drop probability is computed from the bit error probability and the bit encoding length of the triplet. By modeling the encoding and modulation schemes of semantic triplets alongside the channel model, the bit error probability for each user can be obtained. The bit error probability has a complex exponential coupling relationship with transmit power allocation decision \cite{ref7}. It is noteworthy that the semantic triplet drop probability is influenced by the transmit power allocated to the triplet and wireless channel model.

The quality of image semantic transmission has a complex coupling relationship with semantic importance and semantic triplet drop probability. The image transmission quality is defined as $\sum_{j=1}^NI_j\times\left(1-P_{d_j}\right)$ , where $N$ represents the number of extracted image triplets, $P_{d_j}$ is the drop probability of semantic triplet $j$ and $I_j$ is the semantic importance of semantic triplet $j$. With the surge in data traffic, the total power consumption of large-model-driven networks becomes a critical issue to consider. Clearly, allocating more power to semantic triplets with higher importance reduces their loss probability. However, under the constraint of limited total transmit power, this also means the less power is allocated to the remaining semantic triplets, leading to semantic loss. Therefore, an appropriate power allocation scheme should be established to avoid high transmission costs and low transmission quality. This resource allocation scheme increases the loss probability of the most important triplet, but it maximizes the overall transmission quality of image data.

\subsection{Diffusion-Model-Driven Resource Allocation}
After the edge server obtains the semantic triplets extracted by the large model, a straightforward approach is to allocate power equally to each semantic triplet for transmission. However, it is energy-intensive and inefficient, according to the above analysis of semantic transmission quality in large-model-driven networks. To meet the growing demands of intelligent networks, a resource allocation scheme that fully leverages semantic transmission quality must be designed. The purpose of resource allocation is to maximize the communication quality of large-model-driven intelligent networks while considering limited transmit power. The diffusion model, as a generative large model, demonstrates significant advantages in the field of wireless network optimization due to its enhanced decision-making capabilities, flexibility and simplicity of implementation characteristics \cite{ref11}. In particular, the iterative nature of denoising network can highly adapt to dynamic wireless environments and accurately obtain optimal resource allocation schemes. In this section, the diffusion model is adopted to efficiently manage the limited power resources of large-model-driven networks through continuous interaction with dynamic wireless environments.

The diffusion model is a generative large model based on a Markov chain. Its core idea is to model the diffusion process of data points in the latent space to learn the intrinsic structure of a given dataset. The model includes a forward process and a reverse diffusion process. The forward process systematically perturbs the data distribution by gradually introducing Gaussian noise to the original data. Subsequently, in reverse diffusion process, by learning a series of denoising steps, the model restores the data distribution, forming a highly flexible and computationally efficient generative model. The ultimate goal of the diffusion model is to train a generative model that can gradually sample from pure Gaussian noise to obtain the optimal decision-making strategy. 

In the power allocation problem, the constraints affecting the optimal power scheme and factors related to transmission quality can be incorporated into the environment. According to the diffusion model, the optimal power allocation decision $\mathbf{x}^{0}$ in the current environment can be gradually disrupted by adding noise until the original features disappear and turn into pure Gaussian noise $\mathbf{x}^{T}$. Subsequently, in the reverse process of probabilistic inference, the optimal power allocation decision generative network $\pi_{\theta}(\bullet)$ can be viewed as starting from pure Gaussian noise and recovering the optimal decision solution based on environmental conditions. When the magnitude of noise added at each step of the forward process is sufficiently small, the reverse diffusion process becomes the posterior probability distribution of the forward diffusion process \cite{ref14}. To achieve incremental sampling from Gaussian noise $\mathbf{x}^{T}$ to obtain real samples, the generative model $\pi_{\theta}(\mathbf{x}_{0:T})$ needs to learn sufficiently good parameters $\theta$ from the training samples.

\begin{figure}
        \centering	
	\includegraphics[width=3.5in]{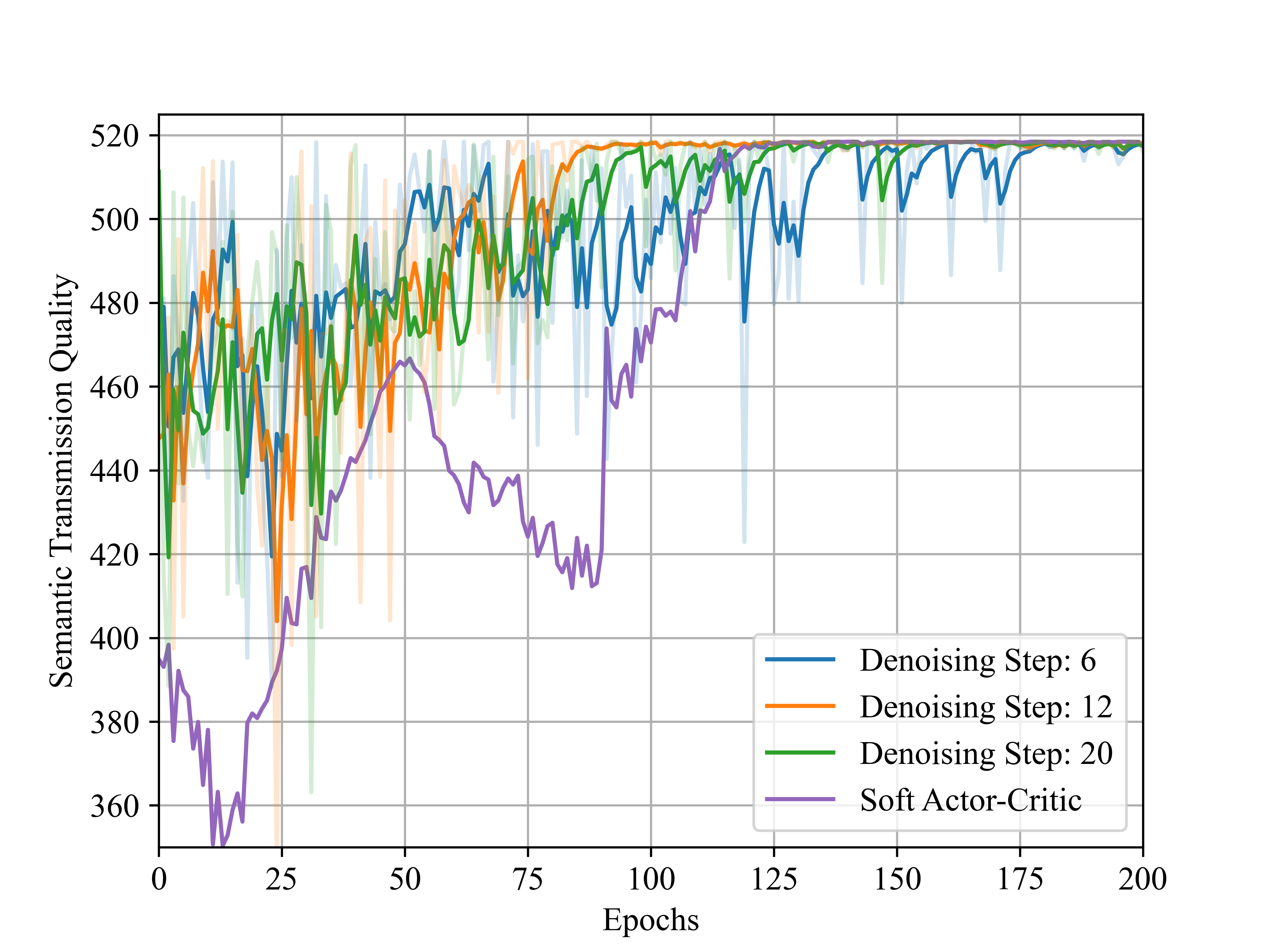}	
	\caption{The semantic transmission quality versus training epochs.}	
	\label{fig4}
\end{figure}

Based on the above analysis, the diffusion process for power allocation is shown in Fig. \ref{fig3}. Specifically, the vector $\mathbf{p}$ represents the power allocation scheme and the vector $\mathbf{e}$ is used to represent the environment, which includes various factors such as the wireless channel model, fixed encoding length of triplets, total transmit power of the BS, and the number of triplets extracted by the RelTR model. In a given environment, our goal is to maximize the expected cumulative reward over a series of time steps, aiming to determine the transmit power allocation factor for each triplet. To achieve this goal, a generative model $\pi_\theta(\mathbf{p}\! \mid\! \mathbf{e})$ is constructed to map the environmental state. The reverse process of the conditional diffusion model can be expressed as follows: initially, a transmit power scheme $\mathbf{p}^{T}$ is sampled from a Gaussian noise sample following the distribution $\mathcal{N}(0,\mathbf{I})$, where $\mathbf{I}$  represents the identity matrix, indicating that the noise is independent across dimensions, and $T$ represents the diffusion step size. Then, iterative sampling is performed according to the parameterized by $\theta$ reverse diffusion chain to gradually infer the optimal power allocation scheme. During the reverse process, each denoising step introduces a new noise source $\varepsilon_{\theta}$, and the training of the denoising process $\pi_\theta(\mathbf{p}\! \mid\! \mathbf{e})$ can be achieved through optimization of $\varepsilon_{\theta}$. Furthermore, a value network $Q_{\nu}$ is used to train $\varepsilon_{\theta}$, which represents the expected cumulative reward of the agent when taking the power allocation scheme and executing it accordingly in the current state. The parameters of the value network are learned and optimized by minimizing the Bellman operator. Ultimately, the optimal power allocation scheme for the large-model-driven semantic network can be obtained.

\begin{figure}
        \centering	
	\includegraphics[width=3.5in]{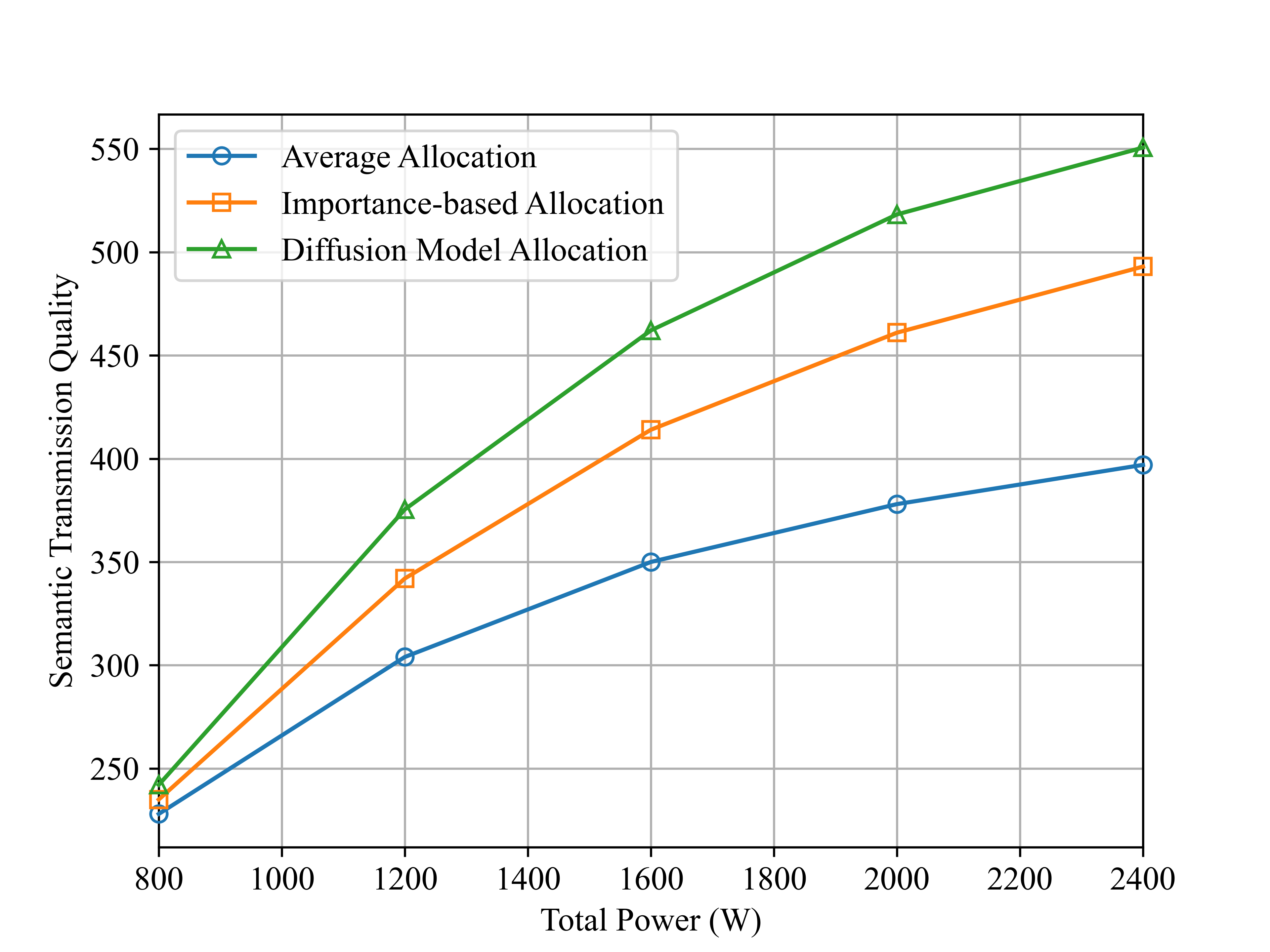}	
	\caption{The semantic transmission quality for diffusion model scheme and benchmark schemes.}	
	\label{fig5}
\end{figure}

\section{Simulation results}
In this section, numerical results are presented to analyze the feasibility of the proposed resource allocation scheme for large-model-driven network. In the considered network, image data are transmitted to users by the BS and the parameters of wireless channel are set as indicated in \cite{ref7}. The visual genome dataset is used to train large-model-driven network. In particular, the BS transmit power is 2000 W in Fig.~\ref{fig4}. Two benchmark schemes are considered, namely, average allocation and importance-based allocation. The former aims to allocate transmit power to each user equally and the triplets are also allocated with equal power. The latter first quantifies the semantic importance of triplets using the CLIP model, and then the importance-based power allocation executes according to the importance weight of each triplet.

 Fig. \ref{fig4} shows the impact of increasing training epochs on semantic transmission quality. The denoising step of diffusion model is set as 6, 12, and 20, respectively. And the deep reinforcement learning method is utilized to compared with diffusion model scheme. It can be seen that the diffusion model of 12 denoising steps exhibits the fastest convergence. The scheme of 6 denoising steps converges slower, which is because the insufficient denoising introduces uncertainty in the generated power allocation scheme and necessitates greater training costs. However, with the increasing of denoising steps, as the denoising steps are 20, the convergence is slower. This could be attributed to excessive denoising, which leads to overfitting during the training of the diffusion model, thereby hindering the adequate exploration of the environment. Therefore, an appropriate number of denoising steps is necessary to achieve a trade-off between adequate denoising and enhanced exploration of the environment by the diffusion model. Although the soft actor-critic \cite{ref15} algorithm and the diffusion model achieve similar results at convergence, the diffusion model allocation scheme with 12 denoising steps converges more quickly. This is due to the diffusion model's more efficient capability to capture complex data distributions.

Fig. \ref{fig5} illustrates the semantic transmission quality comparison between the proposed scheme and two benchmark schemes. It is evident that, as the transmit power increases, the semantic transmission quality of all three schemes improves due to reduced triplet drop probability. Despite the total transmit power varying between 800 W and 2400 W, the proposed power allocation scheme significantly outperforms the two benchmark schemes. This indicates that the proposed scheme thoroughly considers the importance of semantic information and the uncertainties of wireless transmission. Due to the diffusion model's ability to precisely model complex data distributions, the diffusion model allocation scheme achieves better performance. Meanwhile, the decision generation based on diffusion model enhances the flexibility of policy exploration and helps the proposed scheme avoid suboptimal solutions. Moreover, the advantages of the diffusion model in maximizing the exploration of the environment and making resource allocation decisions is demonstrated.

\begin{figure}
        \centering	
	\includegraphics[width=3.5in]{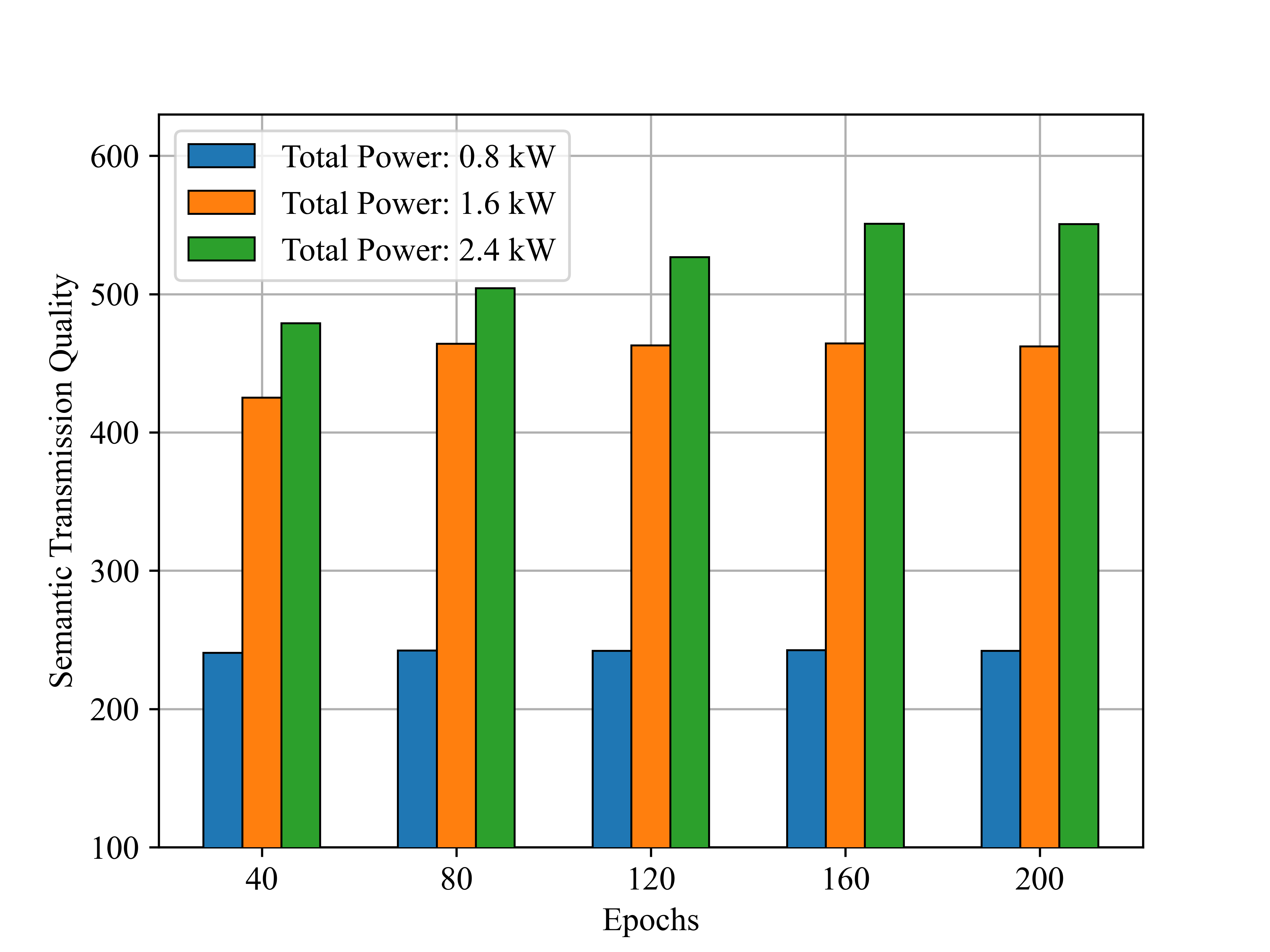}	
	\caption{The semantic transmission quality for diffusion model scheme over different total transmit power constraints.}	
	\label{fig6}
\end{figure}

In Fig. \ref{fig6}, the convergence of the diffusion-model-driven power allocation scheme over different total power constraints is presented. The increase in transmit power makes it more difficult for diffusion model to make decisions. Specifically, the proposed scheme converges at 50 epochs when the total transmit power constraint is 0.8 kW. As the total power increasing, approximately 160 training epochs are required in the case of 2.4 kW power constraint. This is because the increase in total power leads to greater complexity in data distribution, which further causes instability in the exploration of the diffusion model strategy, thereby requiring more training cost to achieve convergence.

\section{Open Issues And Challenges}
Semantic-aware resource allocation scheme plays a significant role in promoting the development of large-model-driven network communication in a positive direction. Nevertheless, there are plenty of inevitable issues and challenges that need further exploration and resolution. Several of them are detailed in this section.

\textbf{Complex Multimodal Scenarios:} Most existing large-model-driven networks typically consider two modalities of data, such as vision and language. For intelligent applications in future networks, such as the metaverse, it is necessary to unify the transmission and intelligent generation of truly multimodal data, including visual images, text, audio, and radar. Therefore, what should be put on the agenda is research on semantic alignment and synchronous efficient transmission of multimodal data based on complex and diverse application scenarios, which demands higher semantic understanding capabilities from large models. The combined research between multimodal large model network design and resource allocation merits further investigation.

\textbf{Adaptive Adjustment Management:} Large models are pre-trained on vast datasets, providing them with good generalization capabilities across various tasks. In future intelligent application scenarios, the surge in personalized user demands and the high precision required for solving specialized problems necessitate further fine-tuning of large-model-driven intelligent mobile networks, resulting in communication delays and computational overhead. Balancing multiple resource needs, including computational cost, parameter count, memory usage, to achieve efficient and adaptive resource allocation for future diverse applications is a critical area for future research.

\textbf{Efficient Network Optimization:} The large-scale and high computational demands make large models difficult to deploy in resource-constrained scenarios. Low-rank and structurally redundant large-model architectures lead to resource waste and reduced application performance in future mobile networks. Model pruning and neural architecture search significantly reduce the need for manual parameter tuning and customized model design, promoting efficient resource optimization in large model networks. How to design optimization schemes for large model networks while developing efficient resource allocation strategies for intelligent networks needs to be further investigated.

\section{Conclusion}
The large-model-driven resource allocation is an essential research direction of future mobile networks. In this article, a novel large-model-driven resource allocation scheme in image semantic-aware network was proposed. The semantic transmission quality model consisting of semantic importance and semantic triplet drop probability was defined as an evaluation metric. To ensure efficient semantic communication of images, the RelTR model was designed to generate image semantic information in textual form and then the importance score was obtained based on CLIP model. Furthermore, the effect of wireless channel transmission was analyzed and a diffusion model for multi-user power allocation scheme to achieve efficient utilization of radio resources was proposed. Simulation results showed that the proposed schemes can relieve communication pressure and enhance the network resource utilization.

\section*{Acknowledgments}
This work is supported in part by the National Natural Science Foundation of China under Grant 62225103, Grant U22B2003, and Grant U2441227, Beijing Natural Science Foundation under Grant L241008, the Fundamental Research Funds for the Central Universities under Grant FRF-TP-22-002C2, Xiaomi Fund of Young Scholar, and the National Key Laboratory of Wireless Communications Foundation under Grant IFN20230201.

\section*{Biographies}
\vspace{-30pt} 
\begin{IEEEbiographynophoto}{Haijun Zhang}
	is currently a Full Professor at University of Science and Technology Beijing, China. He was a Postdoctoral Research Fellow in Department of Electrical and Computer Engineering, the University of British Columbia (UBC), Canada. He serves/served as Track Co-Chair of VTC Fall 2022 and WCNC 2020/2021, Symposium Chair of Globecom’19, TPC Co-Chair of INFOCOM 2018 Workshop on Integrating Edge Computing, Caching, and Offloading in Next Generation Networks, and General Co-Chair ofGameNets’16. He serves/served as an Editor of IEEE Transactions on Communications, and IEEE Transactions on Network Science and Engineering. He received the IEEE CSIM Technical Committee Best Journal Paper Award in 2018, IEEE ComSoc Young Author Best Paper Award in 2017, IEEE ComSoc Asia-Pacific Best Young Researcher Award in 2019. He is a Distinguished Lecturer of IEEE and a Fellow of IEEE.
\end{IEEEbiographynophoto}
\vspace{-30pt} 
\begin{IEEEbiographynophoto}{Jiaxin Ni}
	received her B.S. degree from the School of Computer and Communication Engineering, University of Science and Technology of Beijing, Beijing, China, in 2023, where she is currently pursuing her M.S. degree. Her research interests include semantic communication, large model, and resource allocation in 6G wireless communication.
\end{IEEEbiographynophoto}
\vspace{-30pt} 
\begin{IEEEbiographynophoto}{Zijun Wu}
	received her B.S. degree from the School of Computer and Communication Engineering, University of Science and Technology of Beijing, Beijing, China, in 2021, where she is currently pursuing the Ph.D. degree. Her research interests include IRS, mobile edge computing, and resource allocation in 6G wireless communication.
\end{IEEEbiographynophoto}
\vspace{-30pt} 
\begin{IEEEbiographynophoto}{Xiangnan Liu}
	received his B.S. degree and Ph.D. degree from the School of Computer and Communication Engineering, University of Science and Technology of Beijing, Beijing, China, in 2019 and 2024. He is now a Postdoctoral Researcher with the KTH Royal Institute of Technology, Sweden. His research interests include access control, beamforming, and resource allocation in future mobile communications.
\end{IEEEbiographynophoto}
\vspace{-30pt} 
\begin{IEEEbiographynophoto}{V. C. M. Leung}
	is a Professor of Electrical and Computer Engineering and holder of the TELUS Mobility Research Chair at UBC. He has co-authored more than 1000 technical papers in the area of wireless networks and mobile systems. He is a fellow of the Royal Society of Canada, the Canadian Academy of Engineering, and the Engineering Institute of Canada. He is a winner of the 2017 Canadian Award for Telecommunications Research and the 2017 IEEE ComSoc Fred W. Ellersick Prize.
\end{IEEEbiographynophoto}

  \vfill

\end{document}